\documentclass[12pt]{article}
\usepackage{comment}
\usepackage{verbatim}
\usepackage{graphicx}
\usepackage[labelfont=bf]{caption}
\usepackage{subcaption}
\usepackage{scicite}

\newcommand{\stupid}[1]{#1}
\newcommand{\eg}{\stupid{\textit{e.g.}}}
\newcommand{\ie}{\stupid{\textit{i.e.}}}

\title{Agent-based modeling of the COVID-19 pandemic in Florida}

\author{
Alexander N. Pillai\textsuperscript{1}
\and
Kok Ben Toh\textsuperscript{1,2}
\and
Dianela Perdomo\textsuperscript{1}
\and
Sanjana Bhargava\textsuperscript{1}
\and
Arlin Stoltzfus\textsuperscript{3}
\and
Ira M. Longini, Jr\textsuperscript{4,5}
\and
Carl A. B. Pearson\textsuperscript{6,7}
\and
Thomas J. Hladish\textsuperscript{1,5,*}
\and
\small{}\textsuperscript{1}Department of Biology, University of Florida,\\
\small{}Gainesville, Florida, USA\\
\small{}\textsuperscript{2}Institute of Global Health and Department of Preventive Medicine\\
\small{}Northwestern University, Chicago, IL, USA\\
\small{}\textsuperscript{3}Office of Data and Informatics,\\
\small{}National Institute of Standards and Technology,\\
\small{}Gaithersburg, MD, USA\\
\small{}\textsuperscript{4}Department of Biostatistics, University of Florida,\\
\small{}Gainesville, Florida, USA\\
\small{}\textsuperscript{5}Emerging Pathogens Institute, University of Florida,\\
\small{}Gainesville, Florida, USA\\
\small{}\textsuperscript{6}Department of Infectious Disease Epidemiology,\\
\small{}London School of Hygiene \& Tropical Medicine, London, UK\\
\small{}\textsuperscript{7}South African DSI-NRF Centre of Excellence in\\
\small{}Epidemiological Modelling and Analysis (SACEMA),\\
\small{}Stellenbosch University, Stellenbosch, RSA\\
\small{}*To whom correspondence should be addressed:\\
\small{}tjhladish \{at\} gmail.com
}
\date{}

\begin{document}

\maketitle
\begin{comment}
% link to spreadsheet of model dev events
% https://docs.google.com/spreadsheets/d/1-rRsKzHqNLxbVBX9wUZrFgmoT6G3Nl7F8GDSJG-i9mo/edit?usp=sharing
\end{comment}

\section*{Keywords}
COVID-19, Decision support, Agent-based modeling, Pandemic response

\section{Introduction}
The onset of the COVID-19 pandemic drove a widespread, often uncoordinated effort by research groups to develop mathematical models of SARS-CoV-2 to study its spread and inform control efforts.  The urgent demand for insight at the outset of the pandemic meant early models were typically either simple or repurposed from existing research agendas.  Our group predominantly uses agent-based models (ABMs) to study fine-scale intervention scenarios.  These high-resolution models are large, complex, require extensive empirical data, and are often more detailed than strictly necessary for answering qualitative questions like ``Should we lockdown?''  During the early stages of an extraordinary infectious disease crisis, particularly before clear empirical evidence is available, simpler models are more appropriate. As more detailed empirical evidence becomes available, however, and policy decisions become more nuanced and complex, fine-scale approaches like ours become more useful.

In this manuscript, we discuss how our group navigated this transition as we modeled the pandemic. The role of modelers often included nearly real-time analysis, and the massive undertaking of adapting our tools quickly. We were often playing catch up with a firehose of evidence, while simultaneously struggling to do both academic research and real-time decision support, under conditions conducive to neither.  By reflecting on our experiences of responding to the pandemic and what we learned from these challenges, we can better prepare for future demands.

\section{Context and Early Development} \label{context-early-dev}
\begin{comment}
    Range of model types used
    Translation of dengue model into COVID
    Assembly of synthetic population & parameterization of model
\end{comment}

When we began working on COVID-19 in March 2020, we initially used simple models that could be deployed in minutes.  These included branching process, compartmental, and basic contact network models to address urgent questions for the State of Florida about the prevalence of SARS-CoV-2 based on limited death data, and then to estimate the initial growth and transmission dynamics of SARS-CoV-2.  Early on, we directed these efforts based on the needs of the Florida Department of Health (FDOH), as well as from officials at the University of Florida (UF). We also used these tools to field questions from journalists from local, state, and national media outlets.

The earliest policy concerns were about when and where we should expect challenges with hospital capacity. These transitioned into forecasting the effects of easing business and school closures. When vaccine candidates began to show substantial promise, questions shifted again, towards understanding the impact of vaccination---first to show how important vaccines could be, and later to inform distribution strategies. These questions can be approached through a collection of different models, or a single model can be developed to address most (if not all) of these topics~\cite{nicholson_interoperability_2022}.   As data regarding the pathogen and disease became more available, and we had time to appropriately retool past work, we pivoted from simpler models to using a more detailed ABM as a general purpose tool~\cite{hladish_tjhladishcovid-abm_nodate}.

While the simple models were adequate for the earliest questions, switching to an ABM enabled us to explore detailed facets of SARS-CoV-2 transmission dynamics and testing nuanced control strategies for the state of Florida. While building a single, complex model is more laborious, it can allow for a more natural representation of highly specific interventions (\eg, ``Is to better to limit restaurant capacity, or operating hours?''), and comparisons to idiosyncratic empirical datasets (\eg, serological data collected sporadically from a particular subpopulation).  The detailed population representation in ABMs can capture complex real world relationships (\eg, that families in particular areas have children of particular ages attending specific schools, and that inter-household friendships are often proximity-based) and thus accurately capture important heterogeneities in the population. Rather than developing a new ABM from the ground up, we opted to re-purpose a model we already had extensive experience with, namely our ABM of dengue transmission~\cite{hladish_tjhladishdengue-abm_nodate}.  This transmission model and our approach to parameterizing it had already undergone a decade of development and analysis for a number of research projects~\cite{chao_controlling_2012, hladish_projected_2016, hladish_forecasting_2018, hladish_designing_2020}.

Importantly, this ABM already captured household-level demography, spatial heterogeneities, transmission based on co-localization of infectious and susceptible agents, and a range of infection- and vaccine-based immunity submodels. The obvious first step was to swap the fundamental transmission mechanism from a host-vector cycle to human-to-human.  We added comorbidities, more nuanced infection states, hospitalization and intensive care effects, and death, with time lags sampled from appropriate distributions.  To understand what additional facets should be introduced to the model, it became critical to actively monitor Twitter, news reports, and pre-print servers, in addition to the SARS-CoV-2 literature.  Parameterizing the model also required substantial effort.  We spent much time locating, cleaning, and combining numerous data sources to inform the construction of our synthetic population.  These included datasets related to location coordinates, workplace types, household demographics, comorbidity distribution, and business patronization patterns.  Group meetings increased dramatically both in frequency and duration, as we developed a collective understanding of the pandemic, shared progress, and identified research goals.

Given a more ideal timeline, we would have generalized our ABM to accommodate different diseases and modes of transmission in a modular manner.  Given the realities of the pandemic, we copied the code repository, gutted everything relating to mosquitoes, and set about rapidly adding COVID-specific features.  While expedient (and common), this approach to computational science tends to be short-sighted.  Several innovations in our COVID ABM could be integrated into the dengue model, which would enable, for example, a more sophisticated within-host immunity model. However, because we did not have time to more thoroughly engineer our ABM framework, backporting updates would be a painstaking and error-prone process.  This is an avoidable situation.  Properly supporting and incentivizing scientific software engineering during peacetime would further improve modelers' ability to pivot reliably in an emergency and then recover more quickly as the crisis cooled.

\begin{figure}[h]
    \centering
    \includegraphics[width=\textwidth]{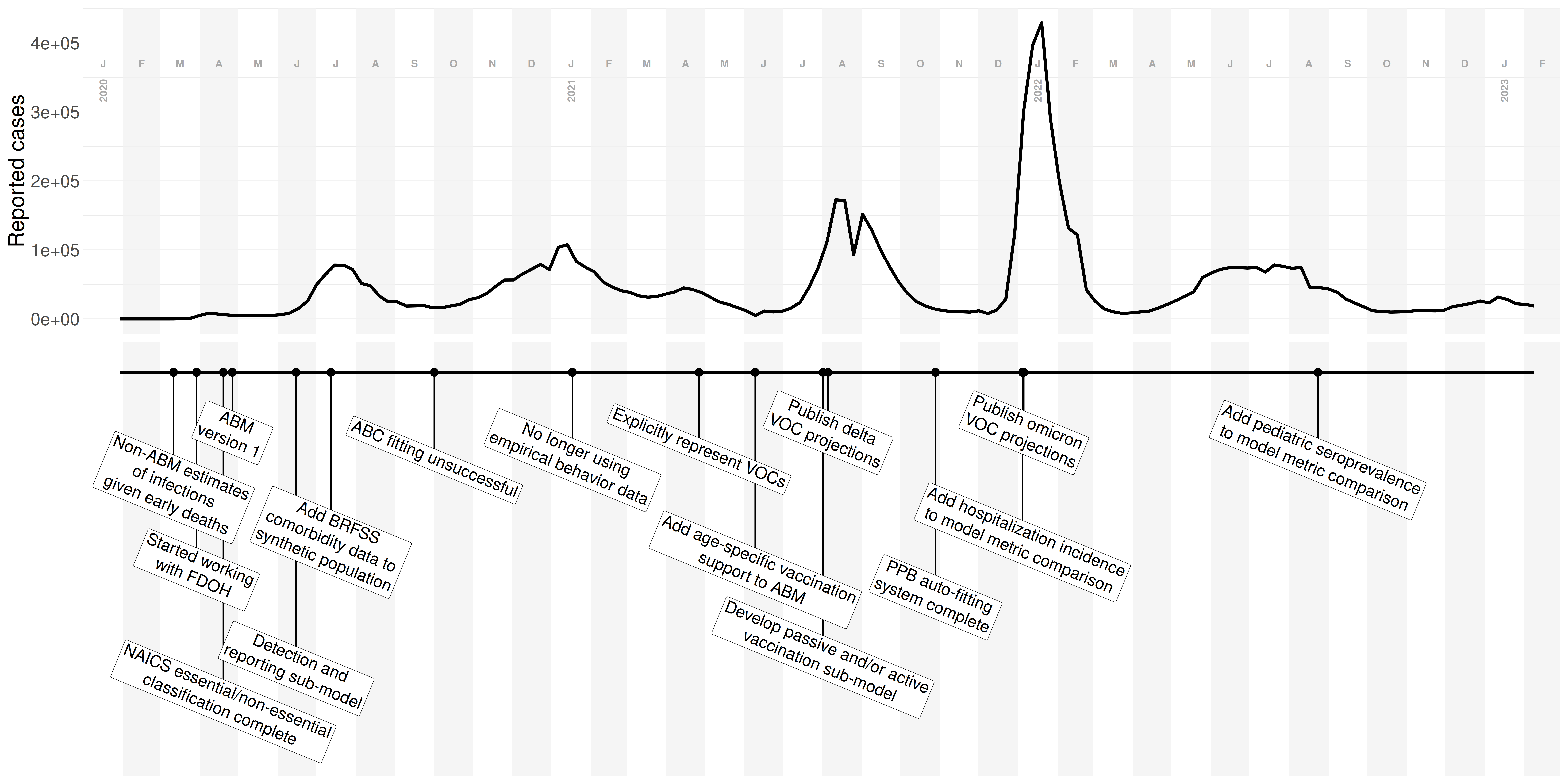}
    \caption{\textbf{Major events in our COVID-19 modeling.}  The timeline (bottom) displays significant milestones of our lab's modeling work, ABM development, and collaborations relative to the time series of COVID-19 reported cases in the state of Florida (top). }
    \label{fig:timeline}
\end{figure}

Our early partnership with FDOH guided how we extended the model to produce answers to specific questions. Conversations with public health officials also helped us better understand resource limitations like testing shortages and hospital capacity. That lead us to incorporate social and logistical limitations influencing transmission and reporting, such as the access to and availability of health resources. We designed systems to represent dynamic case reporting, non-pharmaceutical interventions, behavior, contact patterns, and immunity in order to better model realistic disease dynamics, which we expand upon in later sections (see Figure~\ref{fig:timeline}).

Although simple models are easier to understand, they will tend to not reproduce multifaceted empirical observations.  Simultaneously incorporating many distinct, empirically validated mechanisms and ensuring that macroscopic model outputs match local data gives us greater confidence in projections to inform public health decision-making. In general, such decision support analysis does not allow the feedback available in traditional applications of the scientific method: interventions during an emergency are not run as experiments. With more detailed mechanisms and the reliance on a single set of assumptions, parameters, and data sources, however, we can potentially use other model outputs and more individual level measures (where and when collected) to validate the model~\cite{nicholson_interoperability_2022}.

\section{Model Overview}
We have elaborated technical details of the model previously~\cite{hladish_evaluating_2023, shea_covid-19_2020}. Here, we provide a high level summary of the model, then explore notable features in subsequent sections. As an agent based model, individuals (``agents'') are represented with distinct traits, rather than as counts in various interacting compartments. In our model, those individual traits include, \eg{}, age, sex, COVID-relevant comorbidities and COVID-risk aversion, as well as household membership, social interactions, and daytime activities. These distinguishing traits are the primary generator of heterogeneity in the model (see Section~\ref{sec-heterogeneity}).

Infection risk is based upon the number of co-located infectious individuals, how infectious they are, type of location/nature of interaction, and seasonality. Activity is influenced by both individual decisions (see Section~\ref{behavior}) and top-down interventions that may preclude visiting certain types of locations (see Section~\ref{top-down-interventions}).

Within households, all members interact with each other, and we assume that increasing household size does not dilute the transmission hazard.  In other location types (and for between-household transmission), we assume that risk grows with the proportion of people infected, not the number; the hazard is therefore normalized by the total number of people involved.  We start by assuming the baseline transmission hazard is the same in all location types; by relaxing this assumption, we can, for example, reproduce mortality rates and age-stratified seroprevalence.  More generally, modeling transmission in this way allows our model to explore the application and consequences of location-specific interventions---an advantage that ABMs can offer.

\begin{figure}[h]
    \centering
    \includegraphics[width=\textwidth]{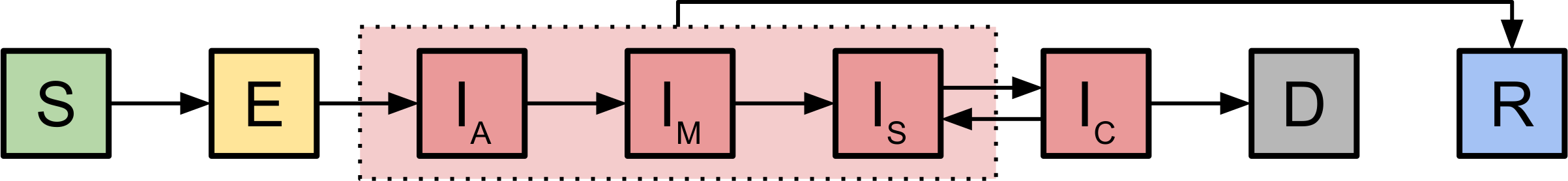}
    \caption{\textbf{Modeled disease state progression.} All individuals begin susceptible ($S$), and upon successful exposure ($E$), will progress to an infectious ($I$) state. Infections can either be asymptomatic ($I_A$) or progress into mild ($I_M$), severe ($I_S$), or critical ($I_C$) symptomatic states. Only critical infections can result in death ($D$); others ultimately lead to recovery ($R$). Recovered individuals have partial immunity against infection that wanes over time, as well as partial protection against future disease that does not wane.}
    \label{fig:states}
\end{figure}

Individuals who experience SARS-CoV-2 infection progress though multiple stages, potentially seeking hospital and/or intensive care. The natural history model proceeds through a series of states (see Figure~\ref{fig:states}) according to empirical waiting time distributions, with probabilistic assignment to various outcomes and treatments. Those probabilities consider individual traits, are modified by vaccination and infection history, and change over time to reflect waning immunity, infecting variant, and changes in medical practice.  In addition to infection history, immunity is affected by innate traits that vary between individuals (see Section~\ref{immunity}).  Individuals that die are removed from the model.

Ultimately, model simulations yield ``observed'' values via a reporting process (see Section~\ref{reporting}), which we compare to available empirical data for both fitting and validation (see Section~\ref{fitting}) and use in decision-support analysis~\cite{shea_covid-19_2020, hladish_updated_2022, hladish_evaluating_2023}.

\section{Heterogeneity}\label{sec-heterogeneity}
\begin{comment}
    Some kinds are low-hanging fruit, i.e. inherent in a data set (household size) or relatively obvious (differences in IFR)
    Why worry about heterogeneity? What empirical data drove these increases in model complexity?
        Very rapid early growth of omicron, but smaller-than-expected overall wave
    What to prioritize, what to ignore, and how to represent heterogeneity
\end{comment}

\begin{figure}[h]
    \centering
    \includegraphics[width=\textwidth]{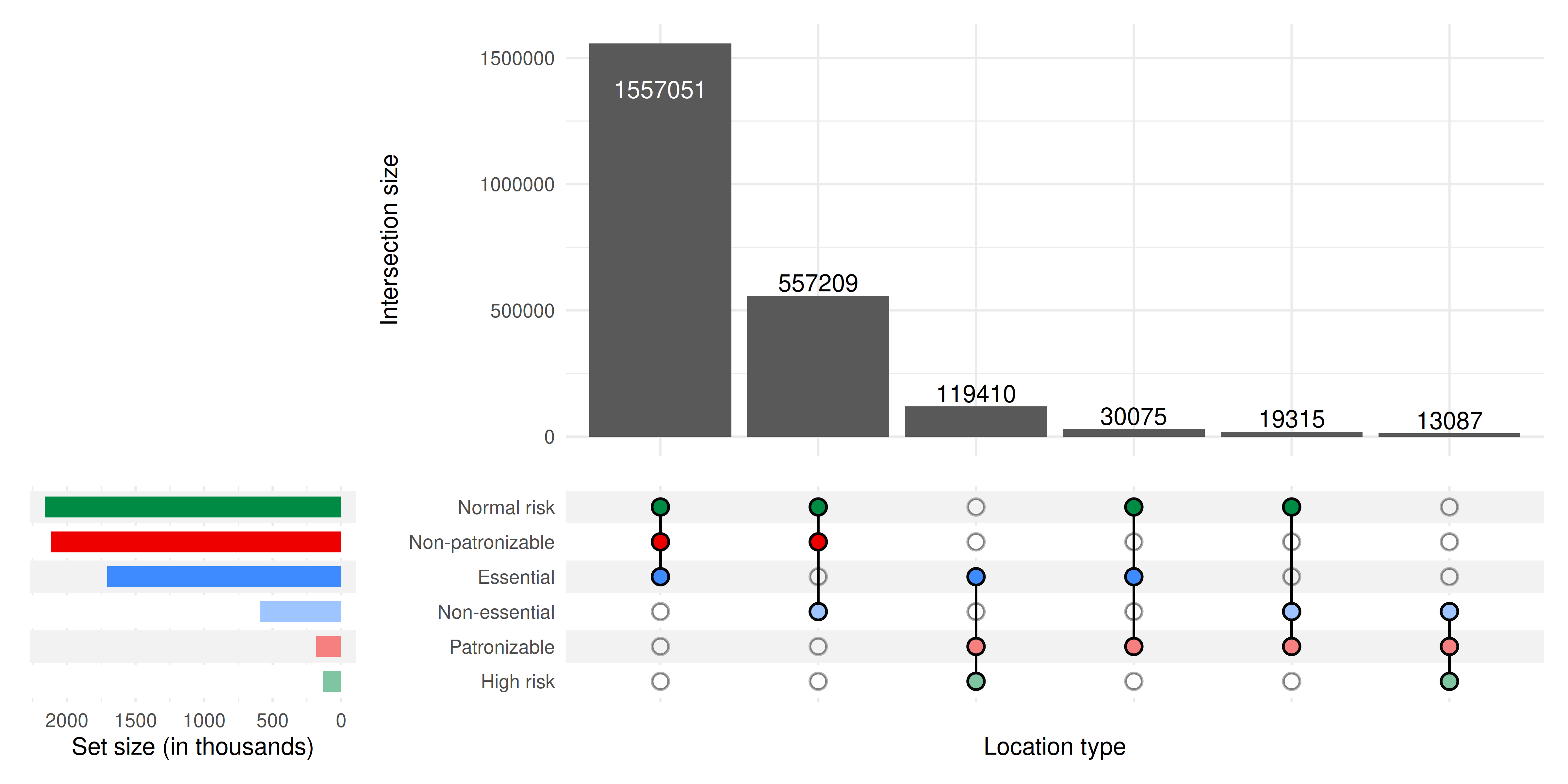}
    \caption{\textbf{Workplaces by type.}  Categorization of workplaces (roughly 2.3M total) in our Florida synthetic population in three dimensions: essential versus non-essential (blue), patronizable versus non-patronizable (red), and normal versus high risk (green). The matrix indicates sets of intersections of these categories and the bar chart above indicates the number of workplaces in that intersection.}
    \label{fig:workplace-categories}
\end{figure}

Many of the policy questions we faced concerned specific groups of people, either for their characteristics (\eg{}, age) or for their activities (\eg{}, where they worked). These groups are not uniformly distributed among the population, but instead tend to mix heterogeneously. Our model captures this sort of mixing by including a range of inherent empirical factors, such as differences in household and workplace sizes, as well as latent traits, such as differences in individual behavior, resource availability, and data reporting.

The sort of heterogeneity arising from population age structure and contact dynamics are largely accounted for by simply creating our synthetic populations based on real-world data as much as possible. That had been a focus of prior work with the precursor dengue ABM; pivoting to COVID-19 thus entailed only minor adaptations to this element of the model.  More effort was required to identify and interpret relevant data sources, particularly for workplaces, schools, and long term care facilities. We used high-precision household samples from the US census to ensure that households in our model accurately mix ages and appropriately represent household demographics, from single-person to large family units~\cite{ruggles_ipums_2020}.  For our model of the entire state (20.5m people), household distributions are specified at the census block group level.  Many research questions can be addressed reliably with a smaller population, however, which means faster simulations and/or more replicates.  In these cases, we used the spatial structure of a single region, typically Marion County, FL (375k people), but sampled households from the census data for the entire state.  We found that temporal dynamics were robust to this simplification, but of course, questions about spatial distribution of burden or deployed resources require the full state model.

By modeling workplaces, nursing homes, and schools explicitly, we can use empirical data to realistically recreate interactions in the population. We obtained the location and North American Industry Classification System (NAICS) business type for the 2.3m workplaces in Florida from National Corporation Directory~\cite{noauthor_national_nodate}.  We use NAICS code to designate businesses as essential or not, patronized or not, and high or normal transmission risk (see Figure~\ref{fig:workplace-categories}), and use data from the Bureau of Labor Statistics to distribute jobs among workplaces~\cite{us_bureau_of_labor_statistics_qcew_nodate}.

Working members of households are probabilistically assigned to those jobs using a gravity model, with the likelihood of matching proportional to remaining spots at a business and inversely proportional to square distance from the person's home. For nursing homes, we use empirical occupancy data and assume facilities follow state laws regarding staffing ratios~\cite{florida_agency_for_health_care_facilityprovider_nodate}.  For schools, we did not find any state-wide enrollment data sets; however, we do have data about school level (primary, K-12, college, etc.), so we assign pertinent-age children and young adults to the nearest schools using the gravity model.

In terms of the latent parameters in the model, households also have a tendency to preferentially connect to households based on similarity in COVID-19 risk tolerance (see Section~\ref{behavior}). This tends to cause more rapid initial spread during waves, as riskier households are likely infected early and to pass on infection, but transmission slows as it spreads among households that are avoiding discretionary interactions.

\begin{figure}[]
    \centering
    \includegraphics[width=\textwidth]{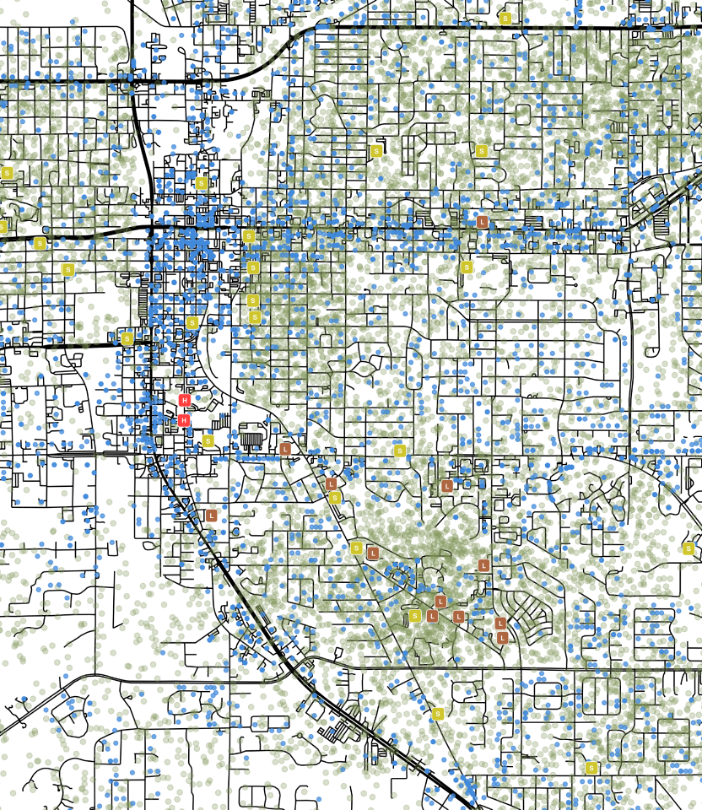}
    \caption{\textbf{Map of synthetic population locations representing Ocala, FL.} Modeled locations show realistic spatial distribution, with workplaces (blue dots) conforming to larger roads (bold black lines) while homes (green dots), schools (green boxes), long-term car facilities (brown boxes), and hospitals (red boxes) cluster around smaller roads (thin black lines).  The ABM does not represent roads \textit{per se}; they are added here for illustrative purposes only. }
    \label{fig:map}
\end{figure}

\section{Behavior}\label{behavior}
We knew first-hand that people's behavior (movement patterns, social interactions, masking, etc.) was changing over time, and in ways that were likely important drivers of transmission.  Although an ABM can potentially represent many different changing behaviors, in the absence of solid empirical evidence about the distinct effects of different types of changes, we focused on changes in social interactions, and later added changes to business patronage.  We observed anecdotally that some people seemed very risk averse, and dramatically modified their behavior, while others sought to modify their activities as little as possible.  Furthermore, people seemed to assort with others who behaved similarly, in families and in friend groups.

This led us to design a behavior submodel, where households were assigned a static risk tolerance value, sampled uniformly on $\left[0,1\right]$.  We then created a household interaction network, where the probability of interacting was correlated with proximity, household size, and similarity of risk tolerance.  By comparing each household's risk tolerance with a time-varying, population-wide societal risk perception (SRP) value, people's behaviors can become more or less restrictive alongside changes in transmission. Specifically, when the SRP value exceeds a household's risk tolerance, its members will avoid patronizing high-risk businesses and will no longer co-localize with households with higher risk tolerance.

Figuring out how to parameterize a new feature is a generic challenge in infectious disease modeling. ABMs generally allow easier translation of new phenomena, but that comes at the cost of complexity in implementation. Like many groups, we initially turned to cellphone mobility data as a proxy for how much time people were spending at home versus at work, school, patronizing businesses, etc. We initially used data from SafeGraph to define the SRP time series~\cite{noauthor_safegraph_nodate}, and found this approach explained epidemiological trends through the summer of 2020. After that, however, the relationships that appeared to previously hold were no longer sufficient to explain trends in transmission, nor were we successful at finding a replacement transformation that did.

We suspect that while mobility changes initially explained most of the variation, individuals and businesses began adopting other measures that were not captured by these data, \eg{}, masking or preferring outdoor activities.  Once those were prevalent, we believe people increased activity and thus their apparent mobility, but without necessarily increasing risk of transmission. To capture these effects, we would need extensive empirical data, which remains difficult to collect and calibrate on a small scale, and likely would fall afoul of privacy protections at the sort of scale necessary to inform a state-wide model. Therefore, we changed the SRP time series to be treated as a latent element of the model and determined during the fitting process (see Section~\ref{fitting} for details). In some sense, the SRP submodel represents the residuals; however, it is constrained by its representation, and in broad strokes the resulting series matches our impression of people's general response to the pandemic.

\section{Top-down interventions} \label{top-down-interventions}
In additional to individual behaviors, the model also supports population-wide interventions. Some of these entail changing the availability of locations for normal activity.

Our model represents transmission as a place-specific process, with different mechanisms and parametrizations depending on interaction setting. Transmission settings include within- and between-households; workplaces; schools; nursing homes; and hospitals (see Figure~\ref{fig:map}).  We simulate the movement of individuals among these kinds of locations, and the overlap of infectious and susceptible individuals in time and place create opportunities for transmission.  In keeping with official Florida policies, we simulate the closure and partial re-opening of schools, in addition to normal closure for weekends and extended holidays.

The State-mandated closure of ``non-essential'' businesses in Spring 2020, and related questions regarding the impact of reopening, led us to incorporate this concept into the model (Section~\ref{sec-heterogeneity}).  All workplaces in the model are tagged with a NAICS business classification; to translate these codes into essential and non-essential categories, we manually categorized all NAICS codes according to the essential industries identified in Florida Executive Order 20-91~\cite{noauthor_state_2020-1}.  This allowed us to evaluate the impact of business closure and reopening directly.

In addition to activity changes, we also represent pharmaceutical interventions. Instead of modeling any specific COVID-19 vaccine product, we generalize the performance of the mRNA vaccines into a single product with a three-dose vaccine series for simplicity. The modeled rollout of vaccines follows the age- and dose-structured distribution of vaccines observed in Florida.  Vaccine efficacy varies by dose and can be diminished by VOC characteristics (\eg, immune escape). Additionally, for those receiving ICU care, we modify the expected mortality probabilities with a logistic interpolation representing the introduction of dexamethasone as a standard component of patient care.

\section{Viral Variants and Individual Immunity}\label{immunity}
As the pandemic progressed, multiple new variants arose and eventually arrived in Florida.  The first of these to spread in Florida, alpha, primarily differed from the earlier strains in its infectiousness, which we initially accommodated simply by gradually changing the baseline transmission hazard in the model.  This allowed us to reproduce the alpha wave, but not as an emergent phenomenon \textit{per se}.  In order to capture more complex variant effects and be able to simulate counterfactuals, we created a new variant submodel, with every infection associated with a variant that is passed infector-to-infectee and could modify infection parameters as needed.

As some of these variants emerged, it was clear that they had different characteristics, and eventually one (omicron) appeared with clear immune-escape properties.  To capture these changes, we have per-variant properties and model individuals with an infection history, potentially with multiple infections. For this particular aspect of our COVID-19 model, its legacy as a dengue model---a pathogen system with multiple strains and complex interaction between infection history and new exposures---was particularly useful.  For work to date, we include four distinct VOCs in the model; in order of arrival, they are wildtype, alpha, delta, and omicron. Each VOC is parameterized with specific infectiousness, pathogenicity, severity, and immune escape values informed from literature and empirical data. The VOCs are introduced by changing the mixture of regular, low probability external introductions. We then observe organic displacement of strains across the entire population driven by differences in VOC force of infection.

In addition to the complex individual immune history arising from variant mixtures, our model also includes vaccination. We assume vaccinations and infections can independently generate immunity in individuals.  With potentially different values by variant, both forms of immunity provide protection against infection, symptoms, severe outcomes, and onward transmission. We allow protection against infection to wane, but the other aspects are held constant. We considered several waning models, including all-or-none and leaky assumptions. We ultimately found that a leaky immunity mechanism, with logistic waning of immune protection over time, was both consistent with the individual-level evidence and able to accurately replicate the variant waves in Florida.

Infection-derived immunity can be thought of as three layers that each can provide protection upon exposure.  During an exposure event, if the exposed individual has an infection history, their infection-derived immunity can generate either (1) short-term broad protection, (2) long-term strain-specific protection, or (3) long-term broad protection.  Short-term protection is defined as a leaky process where where exposures may be resisted by an individually sampled protection level, potentially modified by the variant's immune escape capacity. Long-term, strain-specific protection applies if the individual has had a prior infection with the same strain, and prevents infection either absolutely (if less than 60 days since the previous infection) or probabilistically (85\% against each exposure). Lastly, longer-term broad immunity is drawn statically for each individual, and considered each time a new exposure event occurs.

Vaccine-derived immunity against infection is modeled as individuals having a time-dependent level of protection against infection, waning using the same model as for infection-derived protection. Vaccine protection is also subject to immune escape of the challenging VOC.

\section{Detection and Reporting}\label{reporting}
We rely on empirical time series of reported cases, hospitalizations, and deaths to calibrate our model. However, these data streams are subject to complicated observation, detection, and reporting processes. Reporting rates and delays for both cases and deaths have changed dramatically over time; as a result, a dynamic detection and reporting sub-model was needed to better compare model metrics to their empirical counterparts.

\begin{figure}[]
    \centering
    \includegraphics[width=\textwidth]{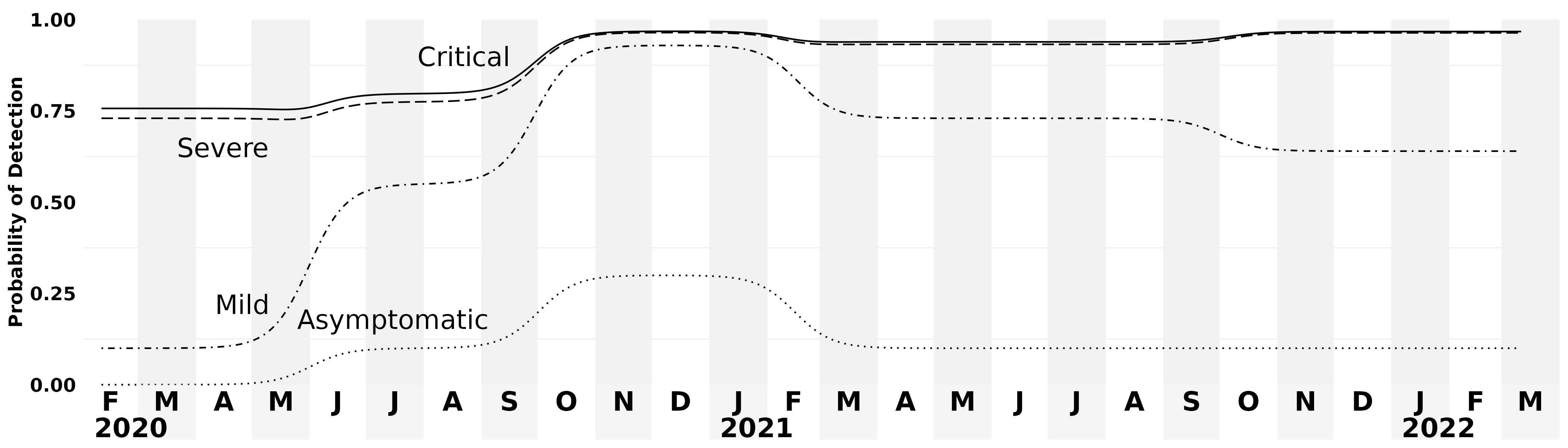}
    \caption{\textbf{Unconditioned probability of detection by peak disease severity.} Detection can occur during any stage of a given infection.  Death detection is not displayed, as we assume perfect death reporting and compare model metrics to total empirical excess deaths.}
    \label{fig:detection}
\end{figure}

We assume that the detection of infections is a function of both disease severity and time. We assume the reporting rate of mild and asymptomatic infections started low ($\textrm{Pr}\{\textrm{detection}\} < 0.125$) and improved over time, until late 2021 when rapid at-home test kits became widely available. While these test kits may have substantially increased individual awareness of infection status, there was no systematic plan by public health agencies to collect those results, likely reducing the official reporting rate of mild and asymptomatic cases (see Figure~\ref{fig:detection}).

Initially our detection and reporting rate assumptions were little more than guesses because of the lack of data. Later, however, we triangulated across multiple data sources to inform the time-varying reporting rate. For example, at several points we used reported cases and deaths in Florida to calculate an observed case fatality ratio (after appropriate time delays), which we compared to global, high confidence estimates to inform the likely reporting rate range at various times. We found that the reporting rate of cases experienced two major improvements over the first two years of the pandemic: the beginning of the 2020 summer wave (which coincides with the expansion of testing criteria) and the beginning of the 2020 fall wave (when asymptomatic testing becomes widely available). Detection and reporting of less-severe infections declined later in the pandemic, however. This first occurred in early 2021, when several mandatory testing programs ended in the state, and then with the omicron wave, as formal testing capacity was exhausted and at-home testing took off.

\begin{figure}[h]
    \centering
    \includegraphics[width=\textwidth]{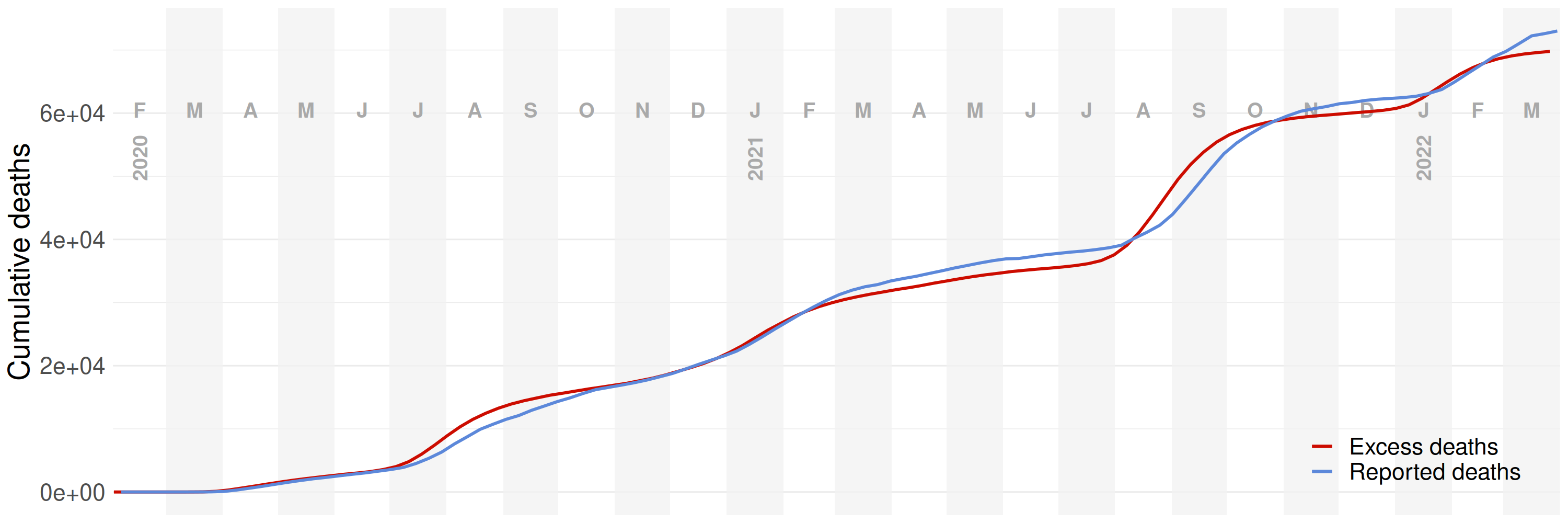}
    \caption{\textbf{Reported vs.~excess COVID-19 deaths in the state of Florida.} Though some inconsistencies exist between these two data streams, we believe that it is a reasonable assumption that all excess deaths in Florida during the simulated time-period can be attributed to COVID-19 deaths~\cite{cdc_trends_2020, noauthor_excess_2023}.}
    \label{fig:death_comp}
\end{figure}

Our data validation efforts changed our preferred source of hospitalization numbers early in the pandemic. Beginning in July 2020, all hospitals were required to report COVID-19 hospitalization to the US Dept. of Health and Human Services (HHS). Reporting to the State was inconsistent by comparison, with a changing subset of hospitals reporting each day.  For all days where hospitalization numbers are available from both HHS and FDOH, substantially more hospitalizations were reported by HHS.  As a result, we include only the HHS dataset in our fitting process. Death data were generally more consistent between sources, with reported COVID-19 deaths agreeing with all-cause excess deaths.  During some epidemic wave peaks (summer waves of 2020 and 2021), however, all-cause excess deaths notably exceeded reported COVID-19 deaths. From conversations with state medical examiners and public health officials, we knew that the process of certifying COVID-19 deaths was complicated and underwent substantial change over the pandemic.  Later, COVID-19 death reporting irregularities were described in detail by an official state audit~\cite{norman_covid-19_2022}.  We ultimately elected to use all-cause excess deaths to avoid these issues (see Figure~\ref{fig:death_comp}). Consistent with that assumption, we set the model’s death reporting rate to 100\% when comparing model outputs with data.

Finally, we modeled the reporting lags for both infections and deaths. They are assumed to follow estimated reporting trends in Florida over the course of the pandemic. We estimated these trends using the linelist of cases and the deaths by day of death time series that was updated daily.

\section{Parameter Fitting Challenges}\label{fitting}

\begin{figure}[h]
    \centering
    \includegraphics[width=\textwidth]{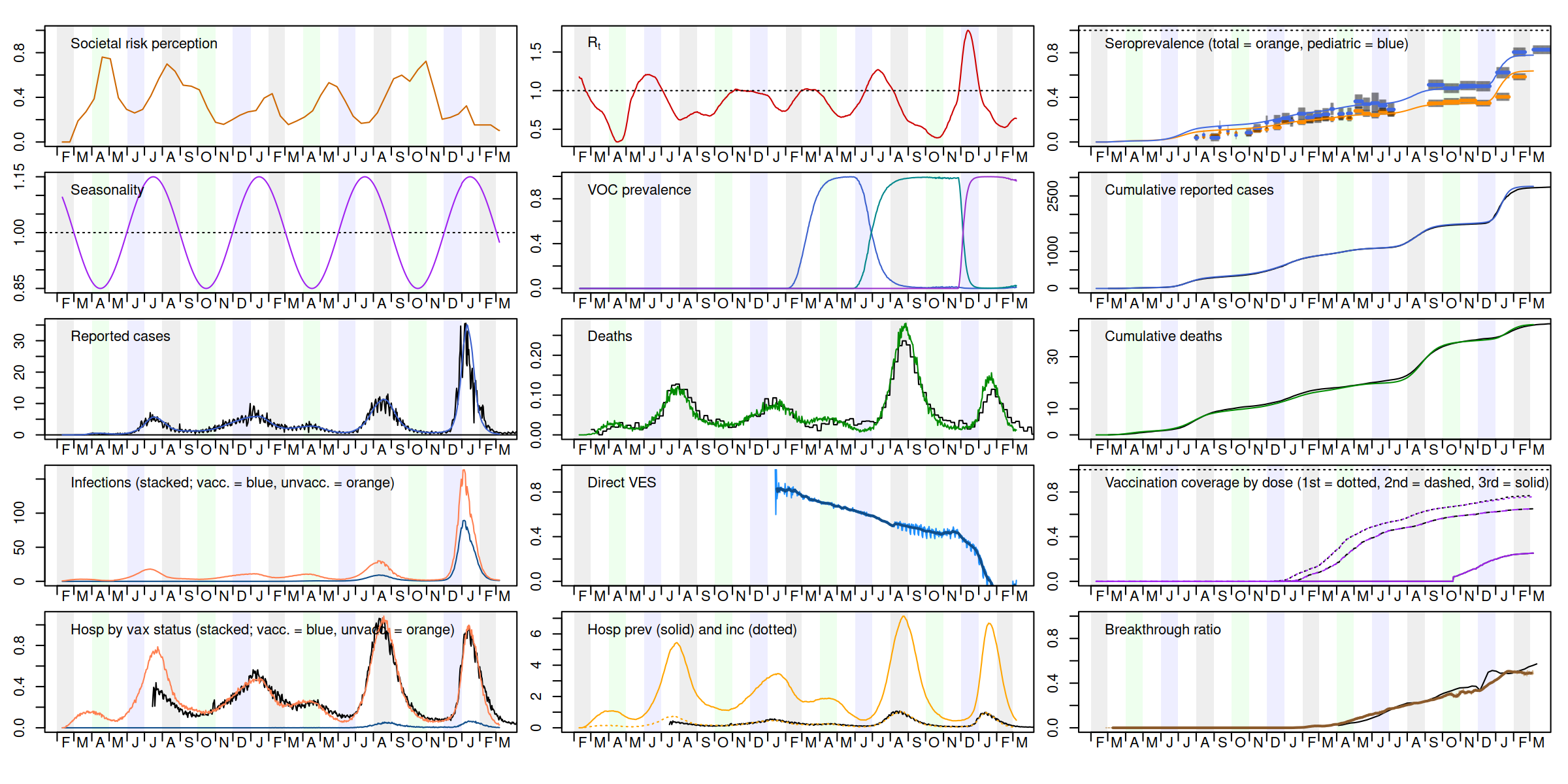}
    \caption{\textbf{Example of our in-house model dashboard.} With this visualization tool, we can easily compare model metrics (plotted in color) to relevant empirical data (plotted in black) during model development and calibration. Vaccination coverage is directly derived from operational data, whereas other metrics emerge as a consequence of the model without being input. }
    \label{fig:dashboard}
\end{figure}

Based on previous success fitting our dengue ABM, which was the ancestor to the COVID-19 model, we initially attempted fitting the new model with Approximate Bayesian Computation by Sequential Monte Carlo (AbcSmc)~\cite{hladish_tjhladishabcsmc_2020}. The first challenge we faced was the size of the new model: with a population ten times larger, and a more complex model, the full state model requires 15--20GB of RAM and a few hours to run.  We addressed this by creating representative, smaller synthetic populations (see Section~\ref{sec-heterogeneity}), and confirmed that important per-capita outcomes were consistent.

Fitting the COVID-19 model presented further challenges, however.  Compared to the dengue model, we had more free parameters, fewer effective measures, and needed more precise fits for our target questions. We found the AbcSmc algorithm was prioritizing easily reproduced metrics (\eg{}, timing of transmission wave peaks), while missing on what we considered more important targets (\eg{}, deaths). Given the urgency at the time, rather than diagnose the fitting problems, we resorted to a much faster \textit{ad hoc} fitting procedure, prioritizing the metrics we believed were most important and reliable, namely deaths, hospitalizations, and seroprevalence. This procedure iteratively switches between manually adjusting free parameters to hit large scale trends and then algorithmically computing the behavior model to precisely reproduce \eg{} wave timing.  While we prefer the objectiveness of methods like AbcSmc when practical, we have confidence in the central estimates of our iterative calibration based on the comprehensive set of metrics reproduced (see Figure~\ref{fig:dashboard}).  We suspect the main downside to our current approach is its inability to properly capture parametric uncertainty and propagate that into forecasts, which is a key feature of Bayesian approaches like AbcSmc.

In the manual stage of fitting, we start with best estimates from the literature, and hand-tune model parameters that are informable based on global evidence but still setting-specific (\eg{}, baseline wildtype transmissibility, reporting probabilities, susceptibility) to align simulated excess deaths and seroprevalence metrics to empirical data.  Once the number of deaths and infections are cumulatively accurate, but perhaps have time series that deviate, we perform behavior fitting.

For the automated stage, we fit the SRP submodel (see Section~\ref{behavior}), though this was also initially a manual process. We would determine the value at regular intervals, then linearly interpolate between these. However, this process rapidly became impractical as the fitted time series grew. The increasing length meant both more points to fit and that it took longer to run the whole simulation. To address this, we developed a greedy fitting algorithm to replace this guess-and-check approach, which enabled us to simulate a sliding window of time, and do the fitting in a single pass.

\begin{figure}[]
    \centering
    \includegraphics[width=\textwidth]{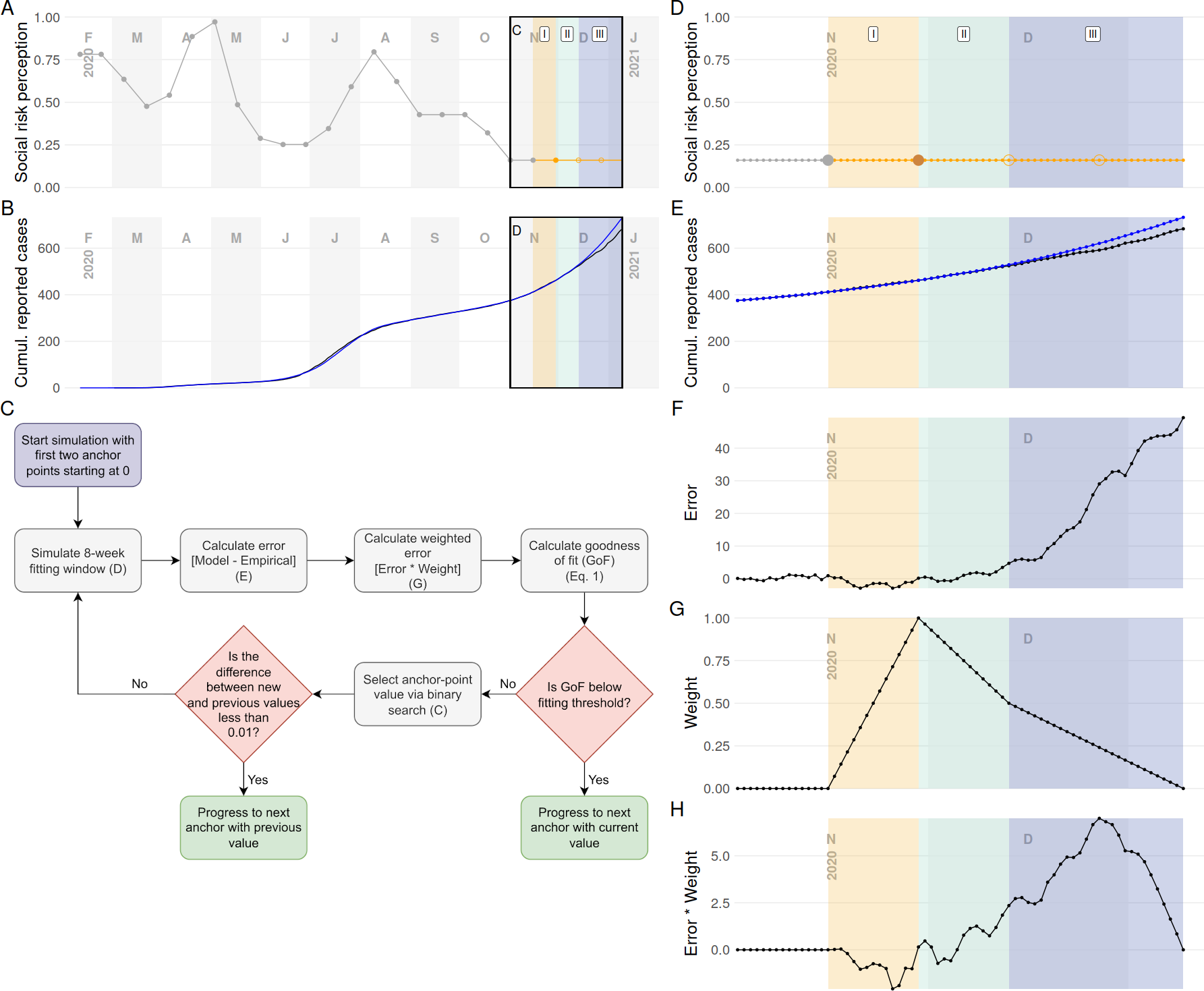}
    \caption{(See next page for caption.)}
    \label{fig:ppb_fitting}
\end{figure}

We perform weighted minimization on the error between simulated and empirical cumulative reported cases (see Figure~\ref{fig:ppb_fitting}), using a iterative propose-check calculation, with binary search to suggest new proposals. During development of this approach, we found that an eight-week sliding window with fitted points every two weeks was the best balance of practicality and accuracy. Because values between fixed SRP points are interpolated, changing a proposed point also changes prior intermediate values, so the window starts with the two weeks prior to the target date.  When evaluating a value for an SRP point, we weight the distance between simulated and empirical cumulative reported cases depending on the day within the entire eight-week fitting window: the most heavily-weighted error occurs on the day of the point being fit and the weight reduces across the time before and after the point. The weighting approach enables us to appropriately balance fitting the target date as well as future dates, which allows this greedy approach to reliably fit SRP in a single pass with only a short look-back period (see Figure~\ref{fig:ppb_fitting}E--H). However, this represented substantial re-engineering of the underlying ABM to add check-pointing---\ie{}, compact storage of the complete state of the model at any particular point.

\clearpage
\addtocounter{figure}{-1}
\begin{figure}[t!]
    \caption{\textbf{Representation of our automated SRP fitting procedure.} This figure illustrates an in-process fitting attempt for a target anchor (solid orange circle). The eight-week fitting window (shaded regions) is divided into three distinct parts: (I in yellow) the two week period prior to the target anchor point, (II in light blue-green) the two week after the target, and (III in light purple) the four subsequent weeks. Region I accounts for the linear interpolation from the previous anchor to the current target. The periods that follow the target point emphasize primarily short term outcomes (Region II) to fit the current anchor, but also incorporate further horizon (Region III) outcomes to ensure later anchors will be reasonably fittable. Panel details: (A) All previously fit (solid gray dots), current (solid orange dot), and future (open orange dot) anchors points. Daily SRP values are linearly interpolated between anchor points (gray and orange lines). (B) Empirical (black) and simulated (blue) cumulative reported cases for the simulated period. For clarity, we zoom into the SRP (D) and cumulative case curves (E) around the current fitting window. For each day, the difference between simulated and empirical cumulative cases is calculated (\ie, error) (F). A weighting function (G) is multiplied by the calculated error to generated a final weighted error curve (H). The weighting favors the day of the anchor being fit and the two weeks before (I) and after (II), but a discounted weight is also applied further out (III). To calculate the goodness of fit for the current anchor value, the weighted error is summed over the fitting window and normalized by a three-day average of empirical cumulative reported cases centered on the current anchor day.}
\end{figure}

\section{Conclusion}
Typically, modeling work entails much more distance between researchers and subject matter. The COVID-19 pandemic was not just an interesting puzzle, or an urgent topic for remote policy-makers: SARS-CoV-2 reshaped our lives, along with everyone else's, as we attempted to grapple with understanding the public health challenge it represented. The experience of these working conditions has highlighted the need to re-evaluate traditional modeling development processes. While conventional arrangements might be suitable for more academic pursuits, they proved detrimental when trying to apply the unique and critical researcher skill set needed to combat this global emergency. 

We were fortunate to have a pre-existing ABM with enough flexibility to accommodate re-purposing to address the pandemic. We were able to reuse an existing vector-borne pathogen model to focus on respiratory transmission, while also substantially expanding that model's structure to accommodate more complex observation processes and surveillance time series, more detailed empirical mechanisms, and new behavioral responses. But to achieve this, we had to invent \textit{ad hoc} processes to collect, skeptically evaluate, and synthesize numerous disparate datasets, including imputing values where data was unreliable or plain unavailable.  At the same time we were trying to elicit sufficiently specified and accurately prioritized questions from stakeholders. We benefited from past attention to software engineering elements during development of the precursor model, enabling these extensions, but these could have been far more thorough. However, even our modest standards were far from ubiquitous in the field, with one of the most prominent models warranting extensive overhauls by professional developers before it could be publicly shared~\cite{adam_special_2020}. Coordinating with public health officials was also a mixed experience: for a few months early in the pandemic, the State of Florida made some effort to incorporate modeling work into decision-making, and until June 2021 provided an enormously valuable data-stream: a publicly available, daily-updated linelist. Globally, such frank bidirectional exchanges proved an essential part of efficacious crisis response. The government-academic collaboration in Florida was unfortunately temporary, however, leading to substantial effort devoted to data interpretation, and having to generalize observations from other settings to ours. Our experiences with public health officials at more local scales was much more durable and productive, although state policies generally prevented local authorities from sharing data.

Our experience with the pandemic suggests a need to emphasize reliability and flexibility in development of modeling tools, rather than on novelty analyses that more readily result in publications. Those modeling tools are more than just the specific models, but the entire tool chain from empirical data to final visualizations, including careful validation of the numerous intermediate data side-products. Traditionally, this sort of work is not rewarded as an academic output, but it has proven critical to applied response and decision-making work. Similarly, the sort of broad base of specialized personnel expertise needed to optimally inform modeling efforts of this sort of complexity rarely comes together to practice this sort of activity. The SMH effort combined the work of roughly 200 researchers on over a dozen models; this pales in comparison to, for example, the 3000 or so authors on the ATLAS experiment at CERN~\cite{the_atlas_collaboration_atlas_2008}. While likely thousands of researchers globally have contributed to analysis of the COVID-19 pandemic, we are unaware of any similar unified effort. Given the relative consequences for those scientific questions, this seems like an area for improvement.

Lastly, relative to public health data, pre-establishing what these streams will look like---\eg, in terms of availability, periodicity, meaning of measurements---would eliminate substantial uncertainty and the work (and rework) necessary to address constantly evolving sources. Each future pandemic will likely manifest differently, but having a baseline to adapt from, rather than inventing standards from scratch over and over, would be substantially more effective. Hopefully, such preparatory work will also provide a bulwark against transient partisan political interests. More open-data approaches will also improve the ability to generalize data to other settings as well as enabling remote groups to contribute to analysis of local outcomes. We found global open-data sources critical to informing our work, but also believe limiting our scope to Florida enabled us to more thoroughly understand the data idiosyncrasies and policy questions in our state.  We expect that even with better data streams, locally-focused groups will remain critical, as effective engagement with policy makers will require local context knowledge and contacts.

\section*{Acknowledgments}
This work was supported in part by The Emerson Collective, The Ron Conway Family, and NIH/NIAID (R56AI148284).  Initial development of the agent based model was supported by the UFII COVID-19 SEED Fund.  CABP was supported in part by the International Decision Support Initiative, which is funded by the Bill and Melinda Gates Foundation (OPP1202541), and by the World Health Organization (2022/1236532).

\section*{Declarations of Interest}
None

\bibliography{references}
\bibliographystyle{Science}

% \section*{Word count}
% \detailtexcount{main}

\end{document}